\documentclass[journal]{IEEEtran}
\usepackage{setspace}
\usepackage{dsfont}
\usepackage{amssymb}
\usepackage{amsmath}
\usepackage{cite}
\usepackage{booktabs}
\usepackage{multicol}
\usepackage{graphicx}
\usepackage{caption}
\usepackage{graphicx}
\usepackage{subfigure}
\usepackage{authblk}
\usepackage{epstopdf}
\usepackage{url}
\title{Resilience of Energy Infrastructure and Services: Modeling, Data Analytics and Metrics}

\author{
\thanks{This paper was submitted.}
\thanks{This paper was prepared in part under the support of the U. S. National Science Foundation under Grants CMMI-1435778 and  ECCS-1549881, and New York State Energy Research Development Authority.}
Chuanyi~Ji,
        Yun~Wei,
        and H.~Vincent~Poor
\thanks{Chuanyi~Ji and Yun~Wei are with the School of Electrical and Computer Engineering, Georgia Institute of Technology, Atlanta,
GA, 30332 USA. E-mail: jichuanyi@gatech.edu, yunwei@gatech.edu.}
\thanks{H.~Vincent~Poor is with the Department of Electrical Engineering, Princeton University, New Jersey, 08544, USA. E-mail: poor@princeton.edu.}
\thanks{Manuscript received  ; revised  .}}

\begin{document}
\maketitle
\begin{abstract}

Large scale power failures induced by severe weather have become frequent and damaging in recent years, causing millions of people to be without electricity service for days. Although the power industry has been battling weather-induced failures for years, it is largely unknown how resilient the energy infrastructure and services really are to severe weather disruptions. What fundamental issues govern the resilience? Can advanced approaches such as modeling and data analytics help industry to go beyond empirical methods? This paper discusses the research to date and open issues related to these questions. The focus is on identifying fundamental challenges and advanced approaches for quantifying resilience. In particular, a first aspect of this problem is how to model large-scale failures, recoveries and impacts, involving the infrastructure, service providers, customers, and weather. A second aspect is how to identify generic vulnerability (i.e., non-resilience) in the infrastructure and services through large-scale data analytics. And, a third is to understand what resilience metrics are needed and how to develop them.

\end{abstract}


\section{Introduction} \label{sec:intro}

Severe weather events such as storms, flooding and extreme temperatures have been occurring across the United States and the world in recent years, increasingly threatening places where large populations and economic activities are heavily concentrated \cite{Hoffman09,Bloomberg13,Sathaye13,WH13}. Among the most affected are the energy infrastructure and services to customers, where weather-induced failures have affected millions of people for days \cite{Doe13,Rudnick11,Bloomberg13}. In response to these disruptions, nation-wide efforts have been initiated on resilience \cite{Rudnick11,Bloomberg13,Nasresilience12,Taskforce12,WH13}. Here resilience refers to the ability to reduce failures under external disruptions and to recover rapidly once failures occur \cite{Bloomberg13,Taskforce12,WH13}.

However, as pointed out by the taskforce report \cite{Taskforce12}, the current understanding of resilience is limited for the power infrastructure under severe weather. It is largely unknown how resilient our infrastructure really is to severe weather \cite{Bloomberg13}. In fact, the problem is not just about fixing the physical infrastructure. Services (i.e., electricity supplies to customers) are pertinent that involve users, service providers (i.e., distribution system operators, or DSOs) and policy makers \cite{Ji16}.

This paper discusses research to date and challenges on resilience. The focus is on how to quantify resilience of the energy infrastructure and services to customers. Here the infrastructure refers to power distribution grids that deliver electricity directly to users. Power distribution grids are found particularly vulnerable to severe weather disruptions, where $90\%$ of failures have occurred \cite{WH13}. Furthermore, the current power distribution infrastructure is not yet fully equipped with the state-of-art technologies for efficient monitoring and protection \cite{Phadke08,Zhao14,Wiseman16}. Services at power distribution grids that involve a large number of customers in disjoint service regions managed by disparate distribution system operators and regulators are particularly challenging. A severe weather event can have a wide geographical span. For example, Super Storm Sandy affected eight million customers in 21 DSO service regions in the United States. In the face of these issues, quantifying resilience remains a challenging problem, involving both infrastructure and services \cite{Bloomberg13,BNLworkshop14,Wei14,Ji16,Zhong05}.

Notably, resilience centers on complex and interacting networks involving weather, the power distribution infrastructure, and a community of customers, service providers and policy makers \cite{Bloomberg13,WH13}. The failure aspect of resilience relates to the interactions between the physical infrastructure and weather. The recovery aspect relates mainly to services. Services depend on complex factors, not only the infrastructure but also DSOs, customers and policies. Both aspects require advanced modeling to incorporate a large number of dependent variables, and data analysis to gain knowledge about what determines resilience. Thus resilience involves a multitude of complex factors from weather to the physical infrastructure, customers, services providers and policy makers. These issues call for not only new thinking and actions from industry but also research that can address fundamental problems underlying the challenges.

We identify three technical challenges within this context. A first challenge is how to model complex interactions among weather, failures at the infrastructure, and recoveries by service providers governed by policies. Mathematical models are needed to describe these important factors, from a large number of local and dependent failures and recoveries, to customer responses and  weather \cite{Ji16,Wei14,Wei13SGC,Wei16}. Meanwhile, it is pertinent to incorporate multiple spatiotemporal scales in these models that span a distribution system locally to service areas regionally \cite{Ji16,Wei14,Wei13SGC}. A second challenge is the development of data analytics that can learn how resilient the infrastructure and services really are from measurements. The power industry has been collecting data on failures and restoration. Such data can potentially be turned into knowledge to guide resilience enhancement. A challenge is the availability of detailed data at a large scale across service territories. Detailed data are owned by DSOs. Large-scale data studies thus call for active participation of DSOs and policy makers. Here, we describe the granularity, scale, paucity and inaccuracy of existing data, from which we learn what new information should be collected. Furthermore, data analytics, although at an early stage for resilience, suggests what knowledge can be learned from the available measurements. Finally, a third challenge is how to measure resilience in a way that incorporates the infrastructure, services, customers and weather. Such resilience metrics are needed for the community to be able to quantify threats and system-wide performance.

\begin{figure}
  \centering
 \includegraphics[width=0.35\textwidth]{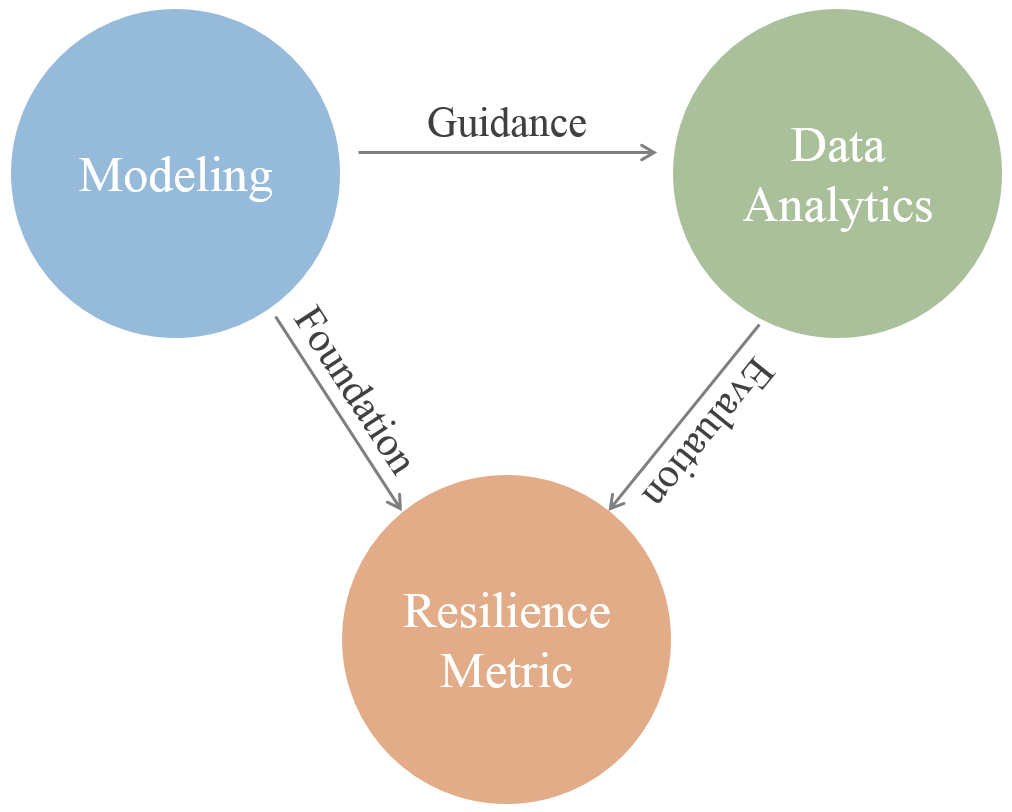} \\
 \caption{Illustration of three inter-related challenges.}
  \label{fig:threeaspects}
\end{figure}

These three aspects are inter-related as illustrated in Fig. \ref{fig:threeaspects}. Modeling lays a foundation to guide data analytics (i.e., on what simple quantities to evaluate from complex failure and recovery processes, and what data to use to gain knowledge on resilience). Meanwhile, modeling provides a basis from which resilience metrics can be derived as system-wide performance. Data analytics provides knowledge and insights on resilience of operational distribution grids and services. Data also underlay measures of resilience metrics and model parameters. Such knowledge should in turn improve modeling and data analytics.

The rest of this paper is organized as follows. The fundamental aspects of the problem and challenges are described in Section \ref{sec:prob}. Modeling is discussed in Section \ref{sec:model}. Data analytics are reviewed in Section \ref{sec:data}.
Resilience metrics are described in Section \ref{sec:resi}. Challenges and open issues are discussed throughout the paper and are summarized in Section \ref{sec:conclu}.

\section{Problem Description and Fundamental Challenges}\label{sec:prob}

The quantification of resilience depends on characterization of the performance of power systems. Resilience can be understood as the ability of the power systems to avoid or reduce power failures and to recover quickly after failure occurrences. These two aspects are inter-related through the concepts of resilience across multiple spatiotemporal scales as stated below.

\subsection{Infrastructure}

A first aspect of resilience is that of reducing failures in the energy infrastructure. Here, as noted above, the infrastructure refers to power distribution systems, the last stage of the grid \cite{Brown08}. Severe weather events (e.g., high winds and flooding) damage power components such as down-wires from fallen debris, damaged transformers or non-functional distribution substations. Component failures in a power distribution system are local (i.e., do not cascade for radial topology) but can involve large numbers of customers and span a wide geographical area \cite{Doe10}. Protective devices, activated by failures or fault currents, are also considered as infrastructural failures since they interrupt electricity supplies to customers \cite{Brown08}. Examples of activated protective devices include open switches and blown fuses \cite{Liu05}. Outages are further caused by failures within a distribution system, where devices downstream lose power but are not damaged \cite{Ji16,Wei13SGC}.

\subsection{Services}

The second aspect of resilience relates to services (i.e., maintaining electricity supplies to customers). Recovery from failures thus signifies the service aspect of resilience. DSOs are responsible for restoring electricity supplies to customers when disruptions occur. Thus, services are provided in a decentralized fashion, where individual DSOs are responsible for their own managed territories. Services are also governed by policies in the form of guidelines from state and federal governments \cite{EEI}. Policy makers also participate actively in recovery processes (i.e., help guide restoration crews as shown in Super Storm Sandy and Hurricane Matthew). Hence customers, DSOs and policy makers form a community relevant to resilient energy services.

\subsection{Multiple Spatiotemporal Scales}

\begin{figure}
  \centering
 \includegraphics[width=0.50\textwidth]{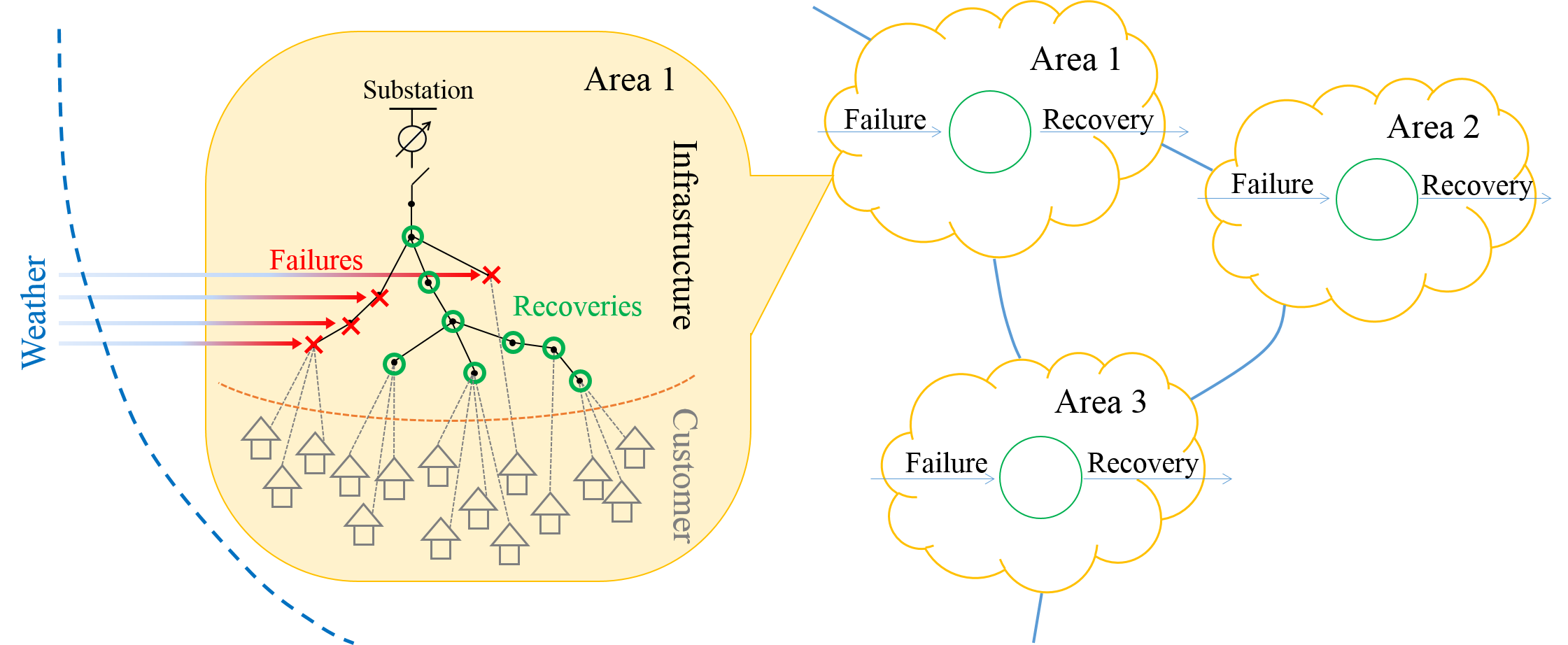} \\
 \caption{Illustration of interactions among weather, infrastructure, and community.}
  \label{fig:coupled_networks}
\end{figure}

Resilience involves interactions among power distribution grids, services to customers, the community and weather as illustrated in Fig. \ref{fig:coupled_networks}. Such interactions occur dynamically across multiple spatiotemporal scales. For example, high winds cause fallen debris that induce failures to overhead power distribution lines in minutes \cite{Wei14}. Outages caused by failures occur in seconds or sub-seconds within a distribution infrastructure \cite{Amin08,Wei13SGC}. Recovery occurs in seconds for restoring outages and in days for difficult manual repairs \cite{Wei14,Ji16}. Spatial scales vary from components in a local distribution system to townships, one service region and multiple service territories \cite{DPS11}.

\subsection{Challenges and Open Issues}

The following challenges emerge:

(a) Modeling: How can we quantify resilience, i.e., starting with modeling, for large-scale weather events? As resilience should be a property of dynamic and dependent networks, models are necessary for integrating local but dependent failures, recoveries, the community and weather at a large scale. Such models are challenging to obtain across multiple spatiotemporal scales for the infrastructure and services jointly (see Section \ref{sec:model} for detailed discussions).

\indent

(b) Data Analysis: How resilient are our power distribution grids and services in the first place?  What should be enhanced for resilience? Data are needed from operational power distribution grids and services to aid in understanding what fundamentally  governs resilience. A challenge is to obtain both detailed and large-scale data from the grid and about weather. Data are owned by individual DSOs governed by policies. Hence data analytics call for collaboration from DSOs and policy makers (see Section \ref{sec:data}).

\indent

(c) Resilience metrics: How can we measure resilience at the network-level involving both the infrastructure and services? Reliability metrics have been used as standards \cite{IEEE1366}. However, these metrics are designed for daily operations rather than severe weather events. Resilience metrics are required to incorporate dynamic characteristics at multiple spatiotemporal scales (see Section \ref{sec:resi}).

\section{Modeling} \label{sec:model}

An objective of modeling is to characterize relationships among a large number of dependent variables. Such variables include weather, failures at the distribution grid level, recoveries, impacts on customers and the community overall. To date, there does not exist such a model that incorporates all these pertinent factors. Different aspects have been studied in prior work, from static models to non-stationary spatiotemporal random processes.

\subsection{Static Models in Machine Learning}

A large body of prior work addresses the modeling of how severe weather induces initiating failures \cite{Angalakudati14,Liu05,Nateghi11,Rudin12}. These models are static, focusing on finding a mapping between weather variables and failures. Such models pioneered the work in this area, starting with one node (e.g., a power distribution line or a component), and one to multiple weather variables \cite{Liu05,Guikema08,Nateghi11}. For example, the failure rate, which is the average number of new failures occurring per mile per hour of overhead power lines, is modeled as a quadratic function of wind intensity in \cite{Brown08}. Fragility, which is the conditional probability of a component failure given weather variables, is modeled as a function of wind intensity or gust, precipitation, and surge elevation respectively in \cite{Reed09a} and \cite{Reed10}.

\begin{figure}
\begin{center}
\includegraphics[width=0.48\textwidth]{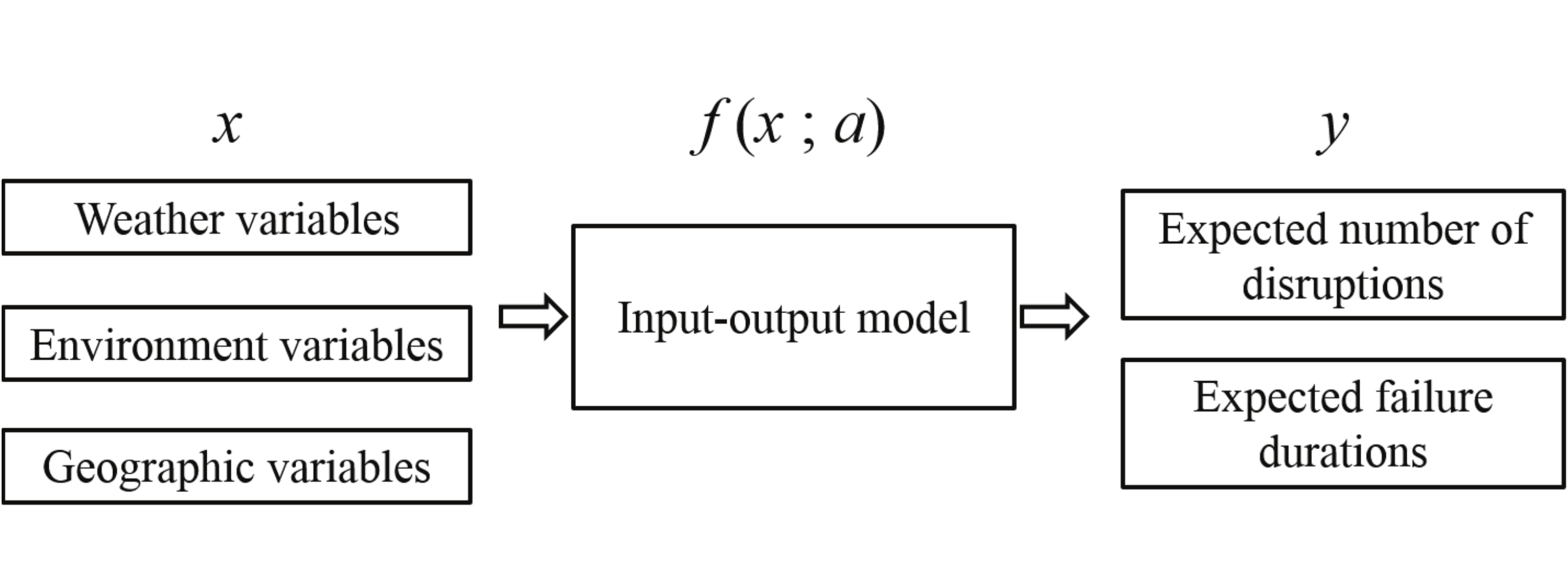}
\caption{Machine Learning View of Static Models.}
\label{fig:stat_model}
\end{center}
\end{figure}

These models, although diverse, can be unified through machine learning as illustrated in Fig. \ref{fig:stat_model}. Consider an $n$-dimensional vector of exogenous variables $x \in R^n$ at a given location for $1 \le n$. Modeling can be viewed as finding a static mapping, $f(x; a): x \to y$, between exogenous variables $x$ and targets $y \in R^m$ for $1 \le m$. Here $y$ describes failures (i.e., as the number of failures or failure durations, or the probability of failures/durations), and $a \in R^l$ is a vector of unknown parameters for $1 \le l$. The data set $D= \{x^{(k)}, y^{(k)}\}$ is obtained on pairs of exogenous and failure variables, where $(x^{(k)}, y^{(k)})$ are the $k$-th sample of the exogenous input $x$ and desired output $y$ of the learning machine. The goal of learning is to obtain either $f$ or the parameter $a$ for a chosen $f$ using data $D$ so that $f$ approximates an underlying mapping from $x$ to $y$. This is clearly a context of supervised learning \cite{Duda}. The models are static, where neither $f$ nor the parameters $a$ nor the inputs and outputs $(x, y)$ vary with time.

The input variables $x$ mainly represent weather, including wind-intensity, speed and gust; as well as precipitation.  Several example models on $f$ have been studied:

(a) Poisson generalized linear model (GLM) \cite{Davidson03}: The number of power failures in a grid cell is modeled as a Poisson random variable with mean $\mu$, where $\ln (\mu) $ is assumed to be a linear function of weather variables $x$.

(b) Negative binomial generalized linear model (NBGLM)\cite{Liu05}: An error term $\epsilon$ is introduced into the GLM to model the dispersion (i.e., the inconsistency between the mean and the variance).

(c) Generalized additive model (GAM) \cite{Guikema08}: The linear function is replaced by a non-linear mapping $f$, including cubic splines and non-parametric models.

(d) Spatially dependent Poisson linear models \cite{Liu08}: Spatial correlation is included in GLM as a multivariate normal distribution across different grid cells.

The above models have been used widely in subsequent works \cite{Han09a,Park12,Winkler10,Cerruti12}. Resulting models identify fallen trees as major causes of power failures \cite{Liu05,Liu07}; and transformers as affected most by severe storms \cite{Davidson03,Han09a}.

Other learning methods have been applied, including classification and regression trees, Bayesian additive regression trees, and multivariate adaptive regression splines \cite{Nateghi11}. Bayesian additive regression trees are found to be most accurate in predicting durations of failures given weather variables \cite{Nateghi11}. Principal Component Analysis (PCA) is found to be effective in learning from correlated weather variables \cite{Han09a}.

The static models assume that failures occur independently of time and locations \cite{Liu05,Liu08}. This assumption on temporal independence is reasonable if evolution of failures is not considered. The assumption on spatial independence can be invalid since locations at sufficiently close proximities may experience similar weather impacts \cite{Liu08}. In addition, certain geo-locations exhibit a higher likelihood of weather-induced failures than the others \cite{Bloomberg13}. Due to these assumptions, certain static models are obtained by aggregating over time and service regions \cite{Liu05}. However, the aggregation may lose spatiotemporal information needed for failure and recovery studies (see Section IV for further discussions).

\subsection{Spatiotemporal Random Processes} \label{sec:model_nonstat}

When sufficiently fine spatial and temporal scales are taken into consideration, failures and recoveries need to be modeled as spatiotemporal random processes \cite{Gallager14,Hajek15,Ji16,Kailath,Wei14}. Such models characterize dynamic interactions of infrastructural failures, services and customers, which are not quantified by static models \cite{Wei14, Ji16}.

\subsubsection{Dynamic Models for Cascading Failures}

The prior works motivate such modeling albeit the problem they consider is on cascading failures that occur at power transmission rather than distribution grids \cite{Dobson07,Bienstock15}. For example, dynamic models are developed for cascading failures through Branching  processes \cite{Dobson04,Dobson07}, Markov decision processes \cite{Wang12}, hybrid system models for random and sporadic failures \cite{Hiskens04}, and other probabilistic temporal models (see \cite{Bernstein14,Hines09,Ilic12} and references therein).

These models are based on stationary probability distributions while weather-induced failure-recovery processes are non-stationary \cite{Wei14,Ji16}. Furthermore, severe-weather induced disruptions span a wide geographical area. Therefore, spatiotemporal processes are needed for weather-induced failures and recoveries.

\subsubsection{Non-Stationary Failure-Recovery-Impact Processes}

Some recent work has developed a spatiotemporal non-stationary model for dependent failure-recovery processes \cite{Wei14,Wei13SGC}. This model is motivated by non-stationary queues \cite{Bertsimas97}. Such models have been applied to failure-recovery processes from severe weather events \cite{Wei14,Wei16}. However, the queuing model is inapplicable when a finer spatiotemporal scale is considered for impacts on customers by each failure and recovery; and restoration is conducted with priorities for critical customers \cite{Emre15}.

A formulation is then developed from bottom-up, starting with failures at the power distribution infrastructure, incorporating service recovery through failure durations, and impacts on customers \cite{Ji16}. Such models integrate a large number of interdependent variables at the finest spatiotemporal scale.

To consider this model, assume failures are already detected. $I_i^{(d)}(t)$ ($d=f,o$) is an indicator function, representing a failure ($f$) or an outage ($o$) for $I_i^{(d)}(t)=1$; otherwise, $I_i^{(d)}(t)=0$. $i$ includes the type, geo-location and system-location of device $i$. Service is characterized by how rapidly power supply is restored to customers \cite{Bloomberg13}; thus represented by downtime duration $D_i (v)$ for failure or outage $i$ that occurs at time $v$. An indicator function $I[D_i (v) > t-v]$ represents the recovery event, where failure or outage $i$ is not yet recovered at $t$, $0 < v < t$. Finally, the impact to customers evaluated at time $t$ is modeled via a function $G_i(v, t)$ for disruption $i$, which occurs at time $v$ for $v < t$. As a simple example, $G_i(v, t)$ is the customer downtime resulting from failure or outage $i$.

Failures, outages, recoveries and costs are dependent for a given weather event, evolving in time and locations. Incorporating randomness from weather disruptions, the spatiotemporal non-stationary random processes model a collection of dependent infrastructural failures, recoveries and costs as coupled processes:

(a) Failure (and outage): $\{I_i^{(d)}(v), i\in S(v), v > 0 \}$,

(b) Recovery: $\{I[D_k (v)>t-v], k \in \overline{S(t)}, 0<v<t\}$,

(c) Cost: $\{G_j(v, t), j\in S(v), 0<v<t \}$.

Here, $S(v)$ and $\overline S(t)$ consist of nodes in normal operation at time $v$ and disruptions at time $t$, respectively. While such a model starts from the finest spatial scale of individual components and customers, aggregation can be done to an area, a township, one service region and multiple DSO territories as illustrated in Fig. \ref{fig:coupled_networks}.

Quantifying completely the spatiotemporal non-stationary random processes is prohibitive since that requires joint probability distributions at all time epochs. The first moments are used in an initial effort, including the time-varying failure rates and marginal conditional probability of downtime duration given failure occurrence time \cite{Ji16}. These simple quantities guide data analytics in Section \ref{sec:data}.

\subsection{Open Issues and Challenges}

The first open issue emerges on how to characterize two generic properties of resilience (Section \ref{sec:prob}): (a) dependent ``networks'' from the infrastructure to customers and service providers impacted by weather, and (b) non-stationarity across multiple spatiotemporal scales. Different models have characterized different aspects of the problem. There are no models yet that characterize all the generic properties. A potential approach is to combine strengths from the machine learning models and spatiotemporal random processes to incorporate diverse factors in a networked setting.

The second open issue is the complexity of modeling. Models can become prohibitively complex when involving spatiotemporal uncertainty, the physical infrastructure and services \cite{NSF08, Brummitt13}. For example, questions arise whether power flows should be incorporated for studying resilience \cite{Bernstein14, Gan16, Lavaei14, Zhu12}; or whether failures can be assumed as detected already so that failure and recovery processes are on (failed, outaged or normal) states of components \cite{Zhao14, Ji16}. The aforementioned models do not involve power flows, which makes the formulation analytically tractable.  Such models also allow data analytics to be conducted using available measurements (see the next section). However, starting from power flows enables failure detection and grid reconfiguration \cite{Zhao14, Gan16}. A challenge is to identify appropriate granularity in modeling dependent variables based on objectives from weather, the infrastructure, services and impacts on customers.

The third open issue is on the community. While the roles of DSOs and policy makers are embedded in service (and the infrastructure enhancement), the existing approaches have not modeled such influence explicitly.



\section{Data Analytics} \label{sec:data}

Data analytics learn knowledge from measurements on failures, recoveries and weather variables. Knowledge learned helps to answer such questions as how resilient the infrastructure and services really are; and what governs resilience or the lack thereof. Here modeling provides a pertinent role of guiding data analytics (i.e., on what measurements to use and what quantities to estimate).

\subsection{Data}

Data determines significantly what knowledge can be learned about resilience.

\begin{figure}
\begin{center}
\includegraphics[width=0.48\textwidth]{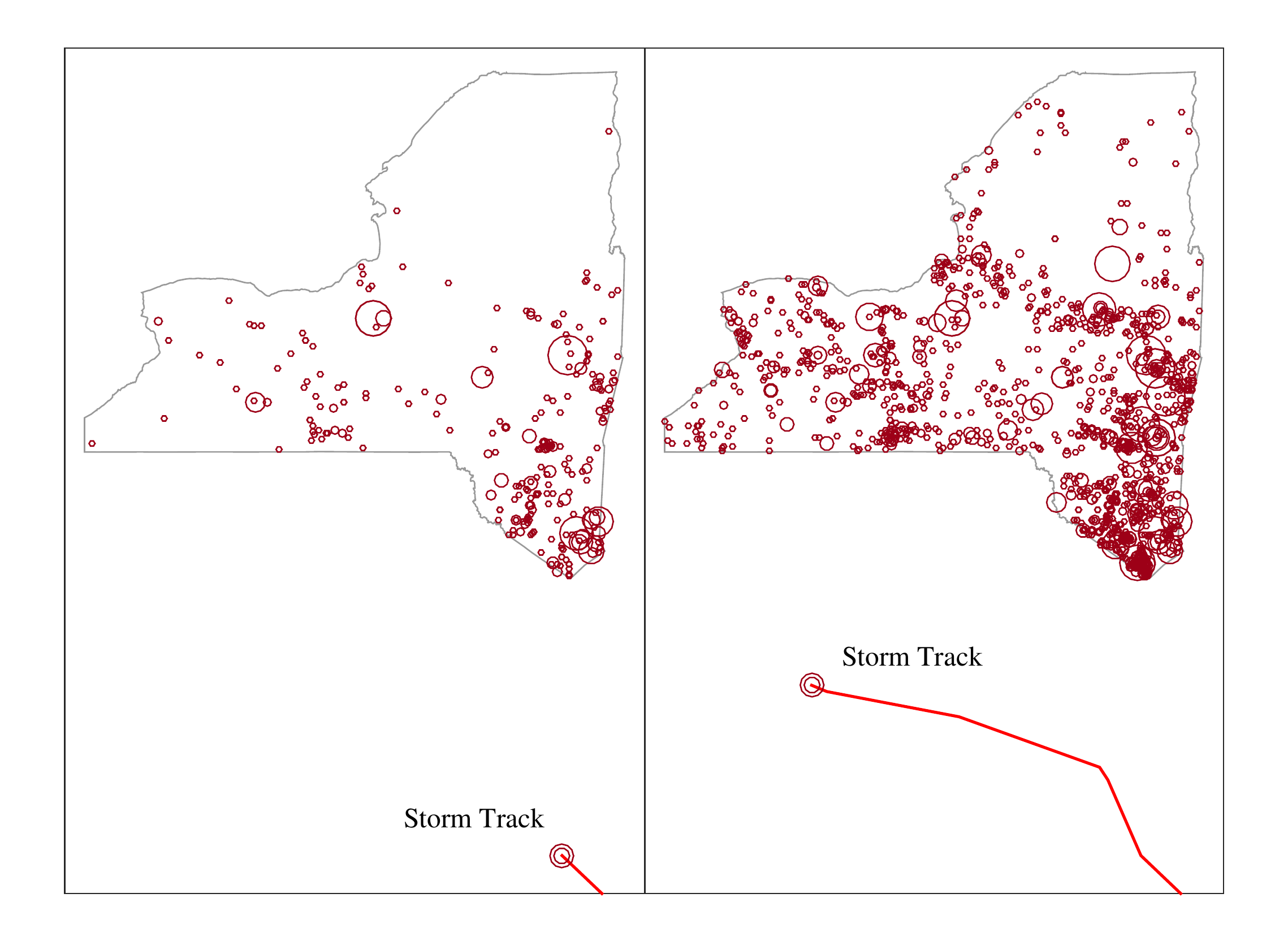}
\caption{Geo-locations and occurrences of failures, and the storm track during Super Storm Sandy.}
\label{fig:sandy_scatter_geo}
\end{center}
\end{figure}

\subsubsection{Data on Failure and Recovery}

The power industry has been collecting data on failures, restoration and impacts to customers \cite{Brown08}. The objectives of this collection have been for outage management, customer communication and reporting \cite{EEI}. Data analysis is yet to become a focus.

As a typical example of granular data, an item on a failure includes ``occurrence- and restoration-time, geo-location, system-location, the type of activated protective device, the number of customers affected'' \cite{Guikema08,Ji16,Wei14}. Here activated protective devices signify the actual failed components, and themselves interrupt electricity service to customers \cite{Davidson03,Han09a}.

One minute is the finest temporal resolution of the available data for failure occurrence and restoration time. Resulting from customer reports on service interruptions, such a time scale is consistent with that of weather-induced failures from dynamic evolution of severe storms \cite{Davidson03, Liu05, Wei14, Ji16}. Failures can further cause outages in a distribution system, where certain components lose power but are not damaged \cite{Wei13SGC}. Outages occur in seconds or less \cite{Amin08, Wei13SGC}. Therefore, the temporal scale of failure data should ideally be less than a second. Such a fine temporal scale is not yet achieved in current data collection. Advanced technologies are required to attain such granularity beyond customer reports. The current resolution on restoration time is also a minute, where failure durations vary from minutes to hours and days.

Geo-locations of failures and outages are provided in exact coordinates of (latitude, longitude) \cite{Ji16}. The location information is usually available on activated protective devices rather than actual failed components \cite{Ji16, Liu08, Wei13SGC}. Data on locations of failed power components are more informative for studying relationships between failures and weather variables \cite{Liu08,Winkler10}. Although not usually available for research, the information on failed devices is known in principle after restoration. Detecting power failures and identifying their locations in real-time have been of research interest, especially with deployment of smart meters, (micro) phasor measurement units (PMUs) for distribution systems, field sensors, and control units \cite{Overbye10, Zhao14, von14, Hahn08, Madingou15}. Overall, accurate geo-locations together with downtime of both failed components and activated protective devices are desirable for studying spatiotemporal variability of severe weather impacts (Fig. \ref{fig:sandy_scatter_geo}).


Accuracy of the data is another pertinent issue. Existing collection methods can fail to generate high resolution data, especially in severely impacted service regions \cite{Coned13}. For example, a large number customer calls in a short time duration hinders failure isolation \cite{Coned13}. Repair crews are typically busy fire-fighting to restore services to customers; data collection on recovery time is thus not a priority \cite{Coned13}. Therefore, automated approaches are pertinent for accurate data collection.

Impacts on customers provide another important source of information on resilience. Available data on the impact is currently measured as the number of customers affected by each failure \cite{Davidson03, Guikema08, Ji16}. Total customer down-time can then be obtained, reflecting impacts from both failures and recoveries \cite{Ji16}.

\subsection{Data on Weather Variables}

Data on weather variables offer pertinent information on external causes of failures and delays on recovery. Commonly-used data on severe storms have been collected on wind intensity and gust, precipitation, moisture, and temperature \cite{Davidson03,Guikema08}. Such data are usually provided by additional sources outside DSO service regions. While an extensive survey of weather data is beyond the scope of this work, well-known example data on wind and precipitation are from the National Oceanic and Atmospheric Administration (NOAA) \cite{NOAA,Davidson03,Han09a,Ji16}. The spatiotemporal granularity of the data varies. For example, the wind speed is measured in minutes and at the centroid of each zip code \cite{Liu07}. The resolution for gust wind-speed is estimated at three-second intervals and each 3.66 km $\times$ 2.44km grid cell \cite{Han09a}. Weather data with a coarse spatiotemporal resolution can be insufficient for terrains with dynamically varying weather conditions. Recent data collection and forecasting techniques improve spatial resolution by incorporating community weather-stations \cite{Deepthunder,NOAA}.

Several DSOs have installed densely distributed weather stations in their service regions, where existing regional weather service is insufficient for dynamic local-conditions. This allows data to be collected on both weather and power failures at comparable spatiotemporal scales \cite{Angalakudati14}. For instance, National Grid has deployed weather stations, each of which covers five square miles in a service area \cite{Angalakudati14}. Central Hudson Electric and Gas has weather stations needed for the varying terrain conditions in the service region \cite{CH16}. San Diego Gas and Electric has sensors and mini weather stations for predicting wildfires and the resulting power failures \cite{SDGE}. Furthermore, a commercial product (Deep Thunder) by IBM offers localized weather prediction at a spatial scale of city blocks\cite{Deepthunder}.

Storm surge and flooding result in damages on power components \cite{Coned13}. However, data on surge and flooding are available at a fewer sources \cite{FEMA}. Synthetic data have been generated from simulation on storm surge and flooding as well as high winds when detailed data are difficult to obtain \cite{Lin12, FZhang15}. A challenge is for simulated data to be sufficiently accurate relating to failures in the infrastructure.

Other exogenous variables include land cover, where residential, forest, commercial, industrial, and transportation land use seem to be related to impacts of service interruptions on customers \cite{Davidson03, Liu05, Liu08}. For example, tree density is an exogenous variable used in prior work that results in down-wire and other initiating failures from high winds \cite{Davidson03,Liu05,Liu08}.

\subsection{Analytics}

As modeling and available measurements lay foundations for data analytics, a pertinent question is what knowledge can be learned from data.

\subsubsection{Outage Management}

DSOs have long been collecting data on failures and recoveries \cite{Brown08}.  A major use of data is on outage management \cite{Coned13,Brown08}. For example, customer reports on service interruption are combined with outage management systems to localize failures. Such information is then used for guiding repair crews. Outage maps are generated for customers on evolution of failures and restoration \cite{CLP_outmap}. Aggregated information on failure and recovery is used for reporting impacts and performance on service restoration \cite{DPS12}. Data analytics have not been a significant part of standard outage management in practice.

\subsubsection{Failure Prediction in One Service Region}

Failure prediction has been studied by the prior works \cite{Angalakudati14,Davidson03,Nateghi11} and \cite{Rudin12}.  One of the first works, although mainly focused on reliability rather than resilience, applies machine learning to predict equipment failures in the  New York City power grid \cite{Rudin12}. An objective is to enable proactive maintenance for reducing severe impacts of power failures resulting in events such as explosions or fires. Data on failures and assets are collected from manholes in multiple years. Reactive point processes are used to learn model parameters from the data \cite{Ertekin15}. Although the power grid in New York City is complex, the data analytics showed promise for failure prediction \cite{Ertekin15, Rudin12}.

Several other prior works pioneered failure prediction using weather data and regression models (see Section \ref{sec:model}). The premise is that if the likelihood of failures can be obtained given weather variables, failures can be predicted through weather forecasts \cite{Angalakudati14,Davidson03,Nateghi11}. Data on failures and weather from one service region are used for parameter estimation and model validation  \cite{Angalakudati14,Han09a,Liu05,Liu07,Reed09a,Reed10,Winkler10,Guikema08,Nateghi11}.

A challenge is that detailed data is often unavailable on failures due to security issues \cite{Winkler10}. As such, the early works have had to use aggregated failure data \cite{Liu05,Han09a,Guikema08,Reed09a,Reed10}. Temporal aggregation results in the number of failures, ranging from one day to an entire period of a hurricane \cite{Christie03,Liu05,Reed09a}. Spatial aggregation of failure locations ranges from a small grid cell of 0.42 km$^2$ to an area specified by a geocode or zip-code \cite{Rourke01,Liu05,Reed09a,Winkler10}.

On other occasions failure and weather data have different granularity \cite{Liu05}. Failure data are then aggregated to match the coarser geographic resolution of measurements on weather and other exogenous variables \cite{Liu05}. Overall, aggregated information over time and locations cannot specify exactly when and where individual failures occur and recover. Thus data on weather and failures, when either are aggregated, can affect the accuracy of a learned model and consequently prediction.

With densely-installed weather stations in a service territory, several recent works have been able to use detailed data on both weather and power failures, resulting in a few failure prediction systems for DSOs \cite{Angalakudati14,SDGE,Deepthunder}.

\begin{figure*}[ht]
\centering
\subfigure[]{%
  \includegraphics[width=0.45\textwidth]{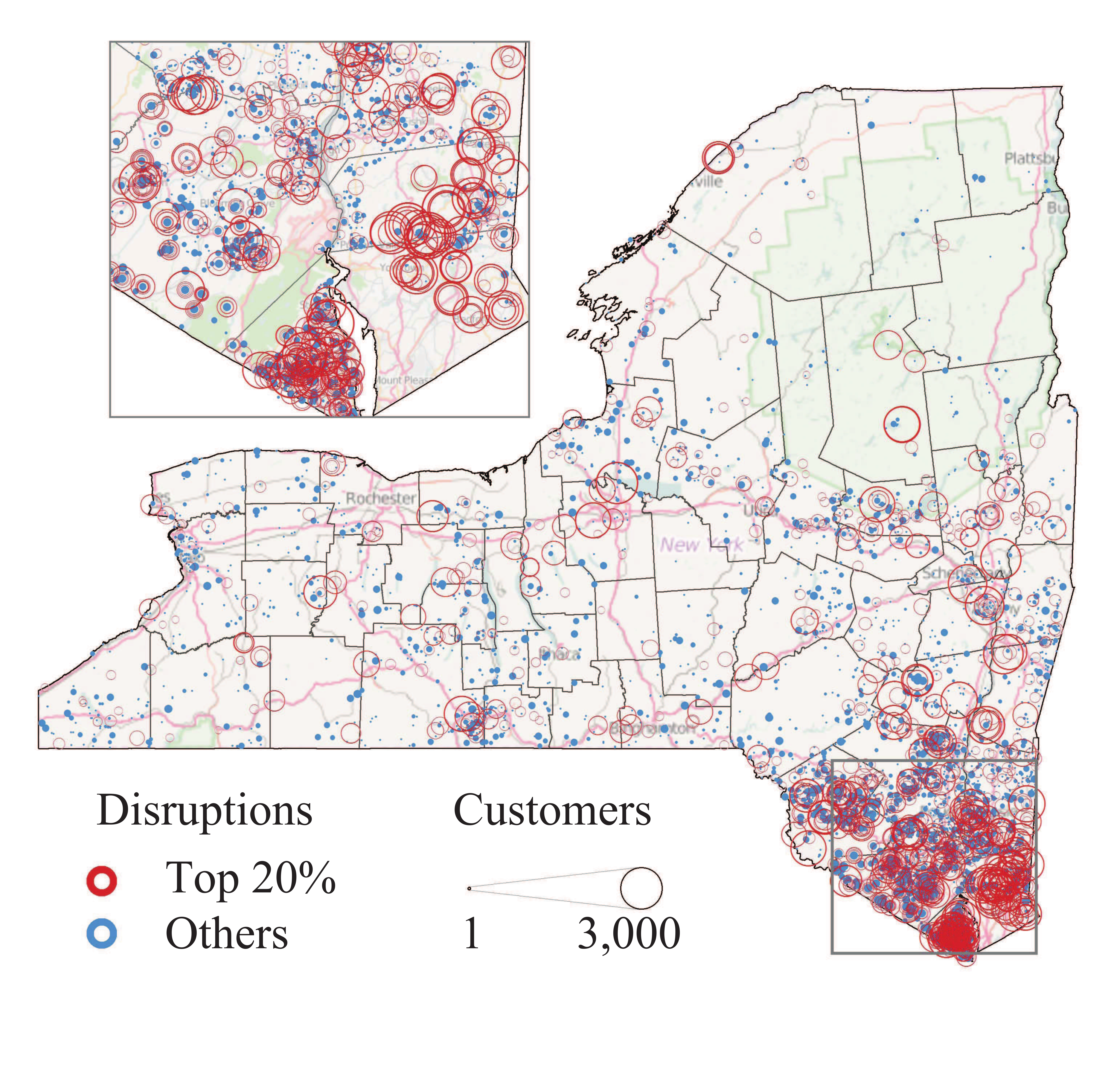}
  \label{fig:2d_scale}}
\subfigure[]{%
  \includegraphics[width=0.45\textwidth]{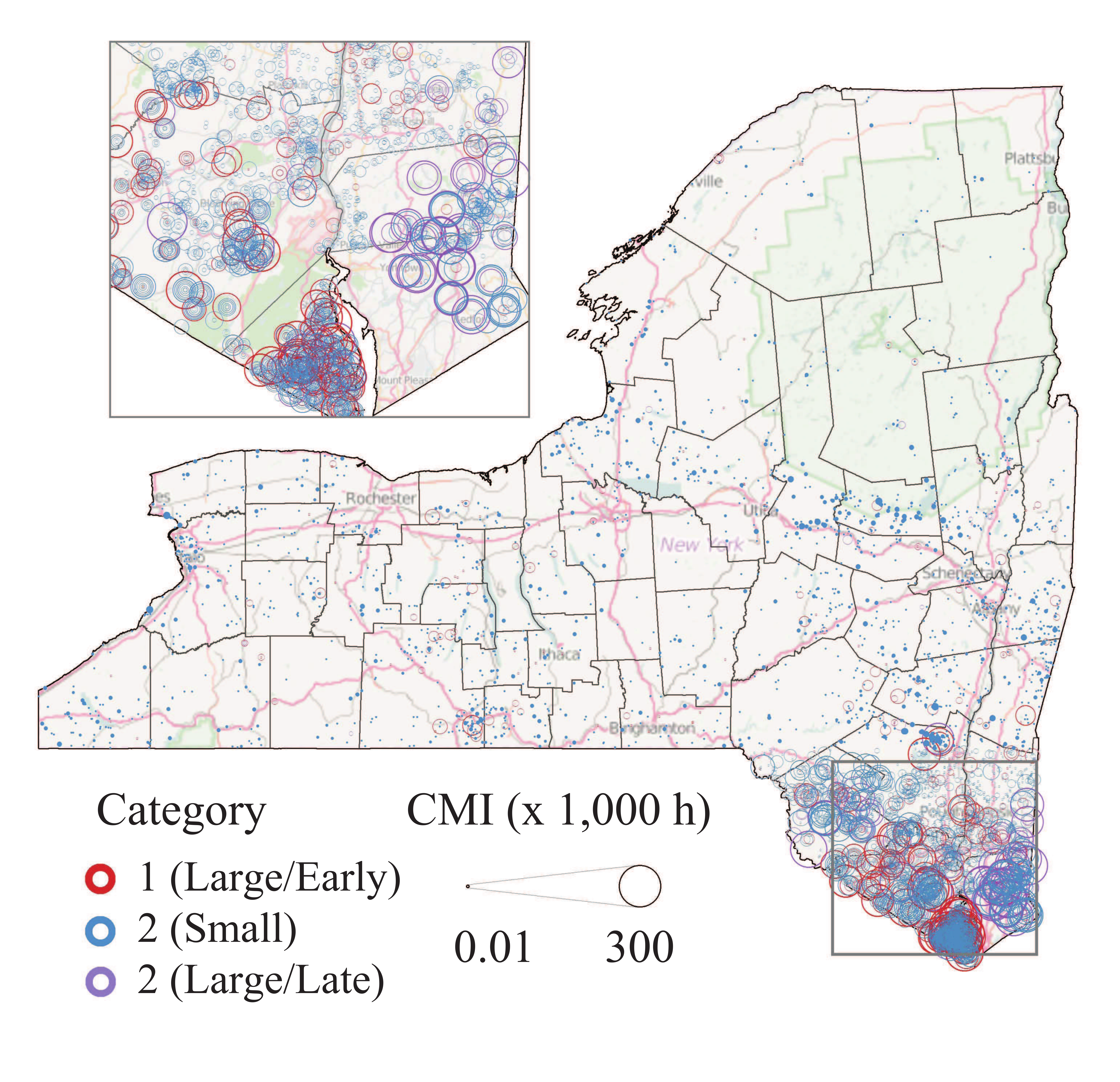}
  \label{fig:2d_cost}}
\caption{Geographical distribution of failures and customer down-time in Upstate New York during Super Storm Sandy. (a) Top 20\% and the remaining failures. (b) Customer interruption hours (CMI), where the colors represent the down-time for large (top percentage) failures that recovered rapidly versus the remaining disruptions. Each marker represents a failure or an outage. Map reproduced using OpenStreetMap and ArcGIS software. The figures are adopted from \cite{Ji16}.}
\label{fig:2d}
\end{figure*}

\subsection{Regression Study at National Scale}

As a severe weather event often spans multiple service regions, a question is how to extend data analysis from one service territory to a regional or national scale. Granular data on power failure and recovery are owned privately by individual DSOs. A recent work explores a novel option of publicly available data \cite{Larsen15}. Such data result from DSO annual report on the IEEE standard reliability indices:
System Average Interruption Frequency Index (SAIFI) and System Average Interruption Duration Index (SAIDI) \cite{IEEE1366}.  SAIFI and SAIDI are the average number of power failures and downtime durations per customer per year. Thus such data are aggregated with a spatial resolution of a service region and temporal scale of a year \cite{DPS12}.

The data are collected at the national scale across the US over the past 13 years \cite{Larsen15}. Data on exogenous variables are also obtained on weather, DSO expenses on reliability, and the density of power lines. The data from all sources are used to learn parameters of regression models \cite{Larsen15}. The failure and duration indices are found to correlate with weather variables, especially when major weather events occurred.

While this approach explores new large-scale data sources, stationarity of the variables may be required so that regression using aggregated data can equivalent to using detailed measurements. Intuitively, the approach is expected to perform well for daily operations when the stationarity is natural for failures and restorations. When including a severe weather event, detailed data are needed at sufficiently fine spatiotemporal scales.


\subsection{Infrastructure and Service Resilience in Multiple Regions}

Owned by individual DSOs, detailed and large-scale data require collaboration from multiple service providers. As a first step, four DSOs and policy makers have collaborated in a recent work, providing detailed data on failure and recovery in Upstate New York \cite{Ji16}. The data span four service regions over $50,000$ square miles that serve two million customers. The granularity of the data is the finest that the current collection can offer \cite{Ji16}. The data consist of failures and recoveries that occurred during Super Storm Sandy and daily operations in 2012. Weather data are not used for the study.

Such detailed data at the large scale enable complementary questions to be studied: whether data analysis can help identify generic vulnerabilities (i.e., non-resilience) in the infrastructure and services. If so, how can data collected by DSOs benefit long-term prediction of resilience through enhancement? A recent work studying these questions is  summarized below \cite{Ji16}.

Guided by the non-stationary spatiotemporal model (Section \ref{sec:model_nonstat}), the analysis of detailed data focuses on a few simple model parameters such as failure rates and conditional probability of downtime durations. The data analysis reveals infrastructural vulnerability illustrated in Fig. \ref{fig:2d_scale}, where local failures, although they do not cascade within power distribution systems, have non-local impacts (i.e., affecting a large number of customers) \cite{Ji16}. A scaling law further characterizes systematically how local failures impact customers: A point of scaling obeys the 20-80 rule \cite{Newman05}, where the top $20\%$ failures affect $~80\%$ of the customers. Importantly, such infrastructural vulnerability was not caused by Super Storm Sandy but exists all along in daily operations. The hierarchical structure of power distribution systems relates to the infrastructural vulnerability, where the majority of the top failures were seen to occur at the higher level of the hierarchy and thus affected a large number of customers. Super Storm Sandy exacerbated such infrastructural vulnerability by increasing the likelihood of such failures. The large scale spanning the four service regions confirms that the vulnerability is common across four DSO service territories.

In contrast, recovery patterns on customer services are different from those failures (Fig. \ref{fig:2d_cost}). A large number (i.e., 89\% of all) small failures that affected the bottom 34\% customers aggregate to 56\% of total down-time. This illustrates challenges for services and requires further study \cite{Ji16}.

\subsection{Challenges and Possible Directions}

The first challenge is how to obtain sufficiently detailed and accurate data. Research to date suggests the necessity of collecting data with sufficient spatiotemporal granularity.  A temporal resolution should be comparable with that of failure and outage occurrences, i.e., at seconds for outages, and minutes for weather-induced failures. Desired spatial resolution corresponds to exact failure and restoration locations. Additionally, large scale (and detailed) data are needed to understand whether knowledge learned is widely applicable. The best available data so far attain accurate spatial resolution, a minute as the time scale, and across several multiple service regions. Aggregated data have been used at the national level. Data acquisition for power recoveries is also insufficient. This is because repair crews have a high priority of restoring services rather than data collection.

Advanced collection systems can help circumvent the data paucity and inaccuracy. A next-generation collection system can potentially gather high resolution data, using widely-deployed micro PMUs in power distribution systems, Advanced Metering Infrastructure and intelligent devices \cite{Karnouskos07, Uddin13,Wiseman16, Zhong05, Hahn08, Meliopoulos10, Peppanen15, von14}. Advanced detection algorithms can then accurately detect and locate failures in real-time to replace customer reporting  \cite{Zhao14}.

\begin{figure*}[ht]
\centering
\subfigure[]{%
  \includegraphics[width=0.45\textwidth]{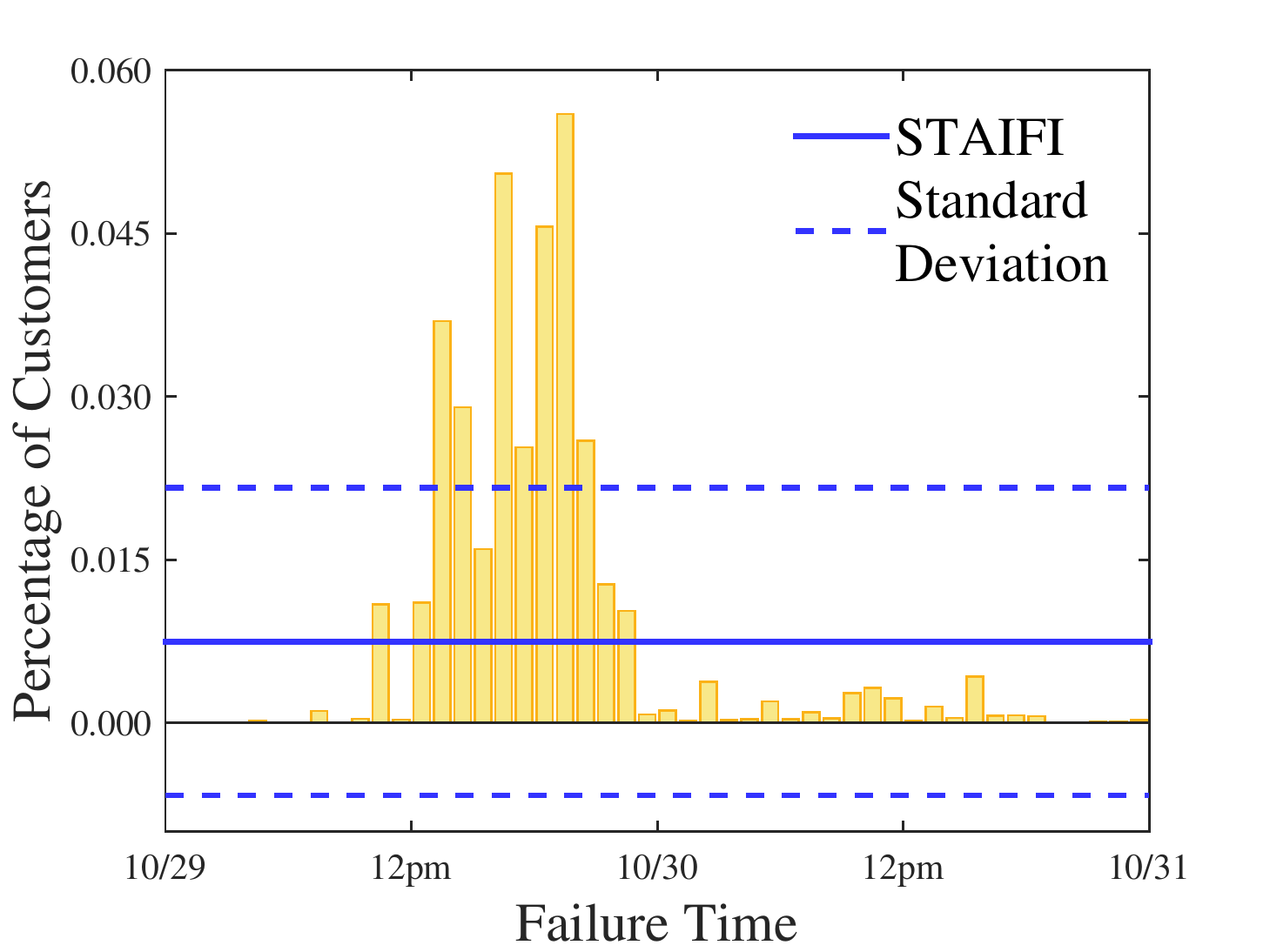}
  \label{fig:staifi}}
\subfigure[]{%
  \includegraphics[width=0.45\textwidth]{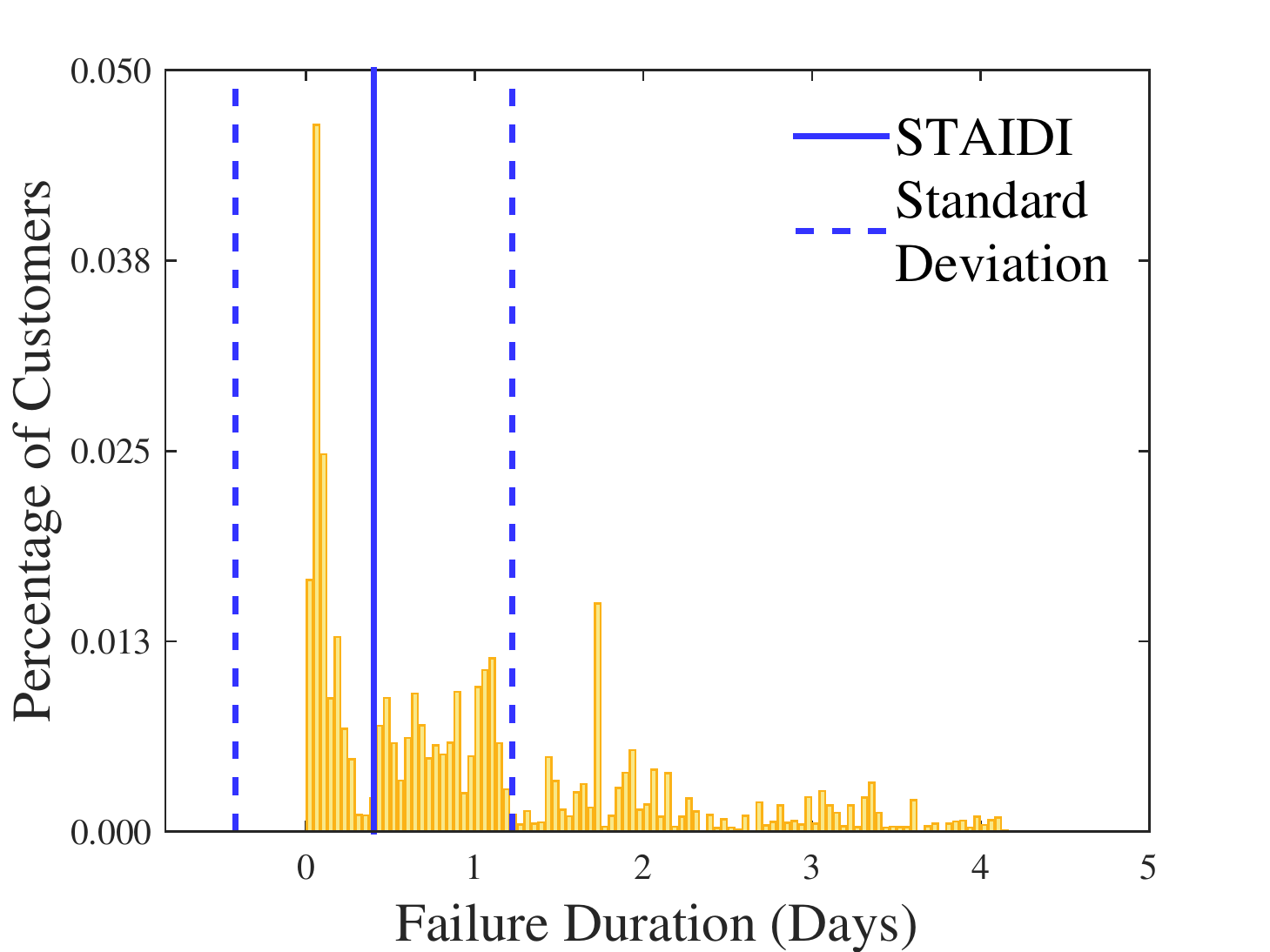}
  \label{fig:staidi}}
\caption{STAIFI and STAIDI from a service region at Upstate New York during Hurricane Sandy. The histograms show the percentage of the affected customers versus (a) time, and (b) interruption duration. The error bars correspond to the standard deviations.}
\label{fig:ieeemetric}
\end{figure*}

\indent

The second challenge is for data on weather and other exogenous variables to have a sufficiently fine spatiotemporal scale. When weather data has a coarser spatial temporal resolution compared to that from the grid \cite{Liu05}, the pertinent and detailed information on infrastructural failures may be wasted \cite{Liu05,Wei14}. An open issue is how to obtain data with sufficient resolution from diverse sources (i.e., the infrastructure, services and weather).

\indent

The third challenge is how to enable collaboration on data analysis. Large-scale and detailed data from multiple DSOs are severely lacking for research due to security and privacy concerns \cite{Bloomberg13}. An important issue is to actively involve DSOs and policy makers to collaborate on data analytics. As shown in the prior study \cite{Ji16}, resilience is for everyone; thus it is possible for DSOs, policy makers and academia to collaborate on data analytics. Procedures that support security and privacy will help enable and expand such collaboration.

\indent

Finally, data analytics, although showing promise of learning knowledge on generic vulnerabilities and predicting failures, are still at an early stage. The potential of data analytics for resilience is yet to be fully explored when measurements become more and more available.



\section{Resilience Metrics} \label{sec:resi}

Modeling and data analytics lay a foundation for resilience metrics. Such metrics are expected to characterize system-wide performance by including all factors from weather to the infrastructure, services and community. Thus resilience metrics need to be derived from models and measured from data. While such metrics are yet to be fully developed, the research to date shows the necessity for new metrics. Open issues include what pertinent variables and how to derive the performance metrics.

\subsection{Invalidity of Standard Metrics}\label{sec:ieeemetric}

Are new metrics needed for resilience in the first place? Two IEEE reliability standards for daily operations (SAIFI and SAIDI) have been extended to severe weather events: Storm Average Interruption Frequency Index (STAIFI) and Storm Average Interruption Duration Index (STAIDI) \cite{Brown97,Reed08}. Given a severe weather event, STAIFI and STAIDI are defined as \cite{Reed08}
\begin{eqnarray}
\text{STAIFI} &= \frac { \text{Total Number of Customers Interrupted}} {\text{Total Number of Customers Served}}, \\
\text{STAIDI} &= \frac {\text{Total Customer Storm Interruption Minutes}} {\text{Total Number of Customers Served}}.
\end{eqnarray}

Now consider a data set from a service region during Super Storm Sandy in Upstate New York. Each data sample is on a failure signified by an activated protective device with the occurrence time, duration and number of customers affected. There are $1334$ failures from October 28, 2012 to October 31, 2012 (see \cite{Ji16} for details). The STAIFI and STAIDI values and their standard deviations are obtained using Equations (1) and (2) and shown in Fig. \ref{fig:ieeemetric}. The standard deviations are so large for both indices that they allow a negative quantity when the error bars are taken into consideration. Such large deviations suggest that STAIFI and STAIDI exhibit too much uncertainty to be valid for characterizing resilience. Therefore,  extending, by brute force, the reliability standards to resilience metrics is not viable.

Having the large standard deviation is not a coincidence but results naturally from non-stationary failure-recovery processes. Non-stationary random processes can of course exhibit time-varying mean functions \cite{Bertsimas97,Ji16,Wei14}. This is clearly shown by the time-varying version of STAIFI and STAIDI from Equations (1) and (2) that are computed using samples at one-hour intervals in Fig. \ref{fig:ieeemetric}. In contrast, the STAIFI and STAIDI are static sample averages by definition, thus insufficient for representing non-stationary failure and recovery processes.




\subsection{Other Metrics and On-Going Studies}

Estimated Time of Restoration (ETR) is another metric used by industry \cite{Coned13}. ETR informs customers of the expected time needed for restoring services after failures. While appealing to users, ETR is difficult to estimate accurately because of the uncertainty and dynamics from non-stationary failure and recovery processes.

Fragility and its variations relate failures to weather variables \cite{Rourke01,Reed08,Winkler10,Yan16}. Such a relationship is necessary to view resilience through potential threats, and thus is promising to characterize a performance metric. A challenge is how to include dynamics and system-wide performance in such a resilience measure.

Dynamic metrics such as Quality and its variations (i.e., Robustness and Rapidity) characterize over time parts of a system or the number of customers in normal operations \cite{Bruneau03, Chang04b, Arghandeh16}. These metrics include dynamic evolution of resilience but are based on pre-assumed functions of time.

Questions arising include, what factors should a resilience metric include; and how to derive such a metric. In principle, a resilience measure should include pertinent factors from weather to failures, recoveries, impacts on customers, DSOs and policy makers \cite{Sandia14}. As resilience quantifies system-wide performance, a fundamental approach is to derive such a metric from bottom-up based on modeling, including weather variables as potential causes, then failures, recoveries, and impacts as consequences \cite{Wei13SGC, Wei16}.


Grounded by the spatiotemporal model, resilience metrics in recent work are derived from the bottom-up to incorporate parts of the factors: non-stationary failure-recovery processes and impacts to customers \cite{Wei13SGC,Wei16}. However, weather and other exogenous variables are not included. A metric $R(t)$ is defined as
\begin{equation}
R(t) = 1 - \frac {1} {C_0} E \Big\{ C(t; d) \Big\},
\label{eq:r}
\end{equation}
where $E \Big\{ C(t; d) \Big\}$ is the expected cost/impact in time $t$ \cite{Wei13SGC,Wei16}. $d>0$ is a threshold on tolerable delays for recovery. $C_0$ is a normalization factor. $R(t)=1$ indicates the best resilience and $R(t)=0$ is non-resilience. $1-R(t)$ is the percentage of cost or impact, evolving with occurrences of failures and recoveries.

The impact/cost has been derived using failure-recovery processes developed through non-stationary queuing models \cite{Wei13SGC,Wei16}. Data from Hurricane Ike has been used to obtain the value of the metric for an operational power distribution grid \cite{Wei13SGC, Wei16}. The failure-recovery-cost processes in Section \ref{sec:model} \cite{Ji16} can potentially be used to evaluate the impact/cost on customers. Thus resilience metrics depend critically on modeling and data.

\subsection{ Challenges and Discussions}

Despite the progress made to date, there are yet to be performance measures that incorporate all three intrinsic characteristics at the system level: Spatiotemporal non-stationary failure-recovery; weather variables; service providers, customers and the community overall. Static metrics such as STAIFI and STAIDI characterize average behaviors of failures and recoveries but not the spatiotemporal evolutions during a severe weather event. The dynamic metrics recently developed include failure-recovery processes and impacts on customers but not weather. The following research questions relating to resilience metrics arise.

(a) What resilience metrics can encompass cohesively the three pertinent characteristics: Exogenous weather variables, spatiotemporal non-stationary failures in the infrastructure, and recoveries of services for customers?

(b) What approaches can lead to such resilience metrics at the system-level, combining weather with failure-recovery-impact processes?

(c) What (additional) data are needed to evaluate resilience of the infrastructure and services?

Answers to those questions are expected to result from both development of system-wide metrics and modeling that incorporates the variables from bottom-up. Extensive data analytics are also needed to obtain values of newly developed metrics and to compare them with the standards.



\section{Conclusion} \label{sec:conclu}

Quantifying resilience of the energy infrastructure and services under severe weather is pertinent but understudied, as shown by the prior works. An immediate need is to understand how resilient the energy infrastructure and services really are. Such understanding enables fundamental enhancement of resilience beyond responding to severe weather. In this context, unique characteristics emerge involving weather, failure and recovery processes in the infrastructure and services, as well as impacts on   community. These characteristics are inter-related and impact resilience together. Therefore, modeling, data analytics and resilience metrics need to be studied cohesively and at a large scale.

Models, when developed for separate aspects of the problem, are found incapable of characterizing these unique properties jointly. Formulating the problem from bottom-up through spatiotemporal random processes has the potential to characterize the interactions of failure-recovery-impact processes. While the modeling framework extends to customers as parts of community, roles of service providers and policy makers have been insufficiently studied. Relationships between weather variables and failures have been studied in a separated context. Models are yet to be developed to incorporate all pertinent factors.

Data analytics, although at an early stage, have started to show promise in learning knowledge about resilience. Data collected by DSOs, when sufficiently detailed, have shown potential in identifying generic vulnerabilities of the infrastructure and services. Large-scale and detailed data are particularly needed from multiple service territories. This provides an opportunity for collaboration among DSOs, policy makers, and researchers. After all, resilience is for the benefit of the entire community. As big data is prospering in many fields of engineering, the resilience problem presents a new application area.

Modeling lays a foundation for deriving resilience metrics. Widely-used IEEE standards are mostly developed for reliability in daily operations. Those metrics, when directly extended to severe weather disruptions, are found to exhibit too much uncertainty to be reliable. Resilience metrics developed from bottom-up can incorporate non-stationary spatiotemporal failure-recovery-impact processes across multiple spatiotemporal scales. However, such metrics do not yet include weather variables. The metrics with weather variables are developed in a separate context. An open issue is how to derive resilience metrics, combining weather with the infrastructure and services.

Full of open issues and challenges, the problem of resilience provides a fertile ground for technical study. Modeling, data analytics and metrics are still very much open for development.  Models and metrics are yet to include cohesively a wide range of exogenous variables. More details from the infrastructure such as power flows may be needed in the study of resilience.

In a broader context, while the methods discussed here focus on the power distribution infrastructure and services under severe weather, the approaches for quantifying resilience can be generalized to other dependent networks in a natural environment.

\appendices

\indent

\section*{Acknowledgment}

The authors would like to thank Ling Wang for help with our literature review on modeling, and Amir H. Afsharinejad, Scott Backhaus, Russel Bent, Steve Church, Timothy Hayes, John Love, Henry Mei, Debasis Mitra, Brian Nugent, Maria Rodriguez, Thomas Spatz, Gregory Stella, Matthew Wallace, Robert Wilcox, Michael Worden and Meng Yue for insightful discussions.

\bibliographystyle{IEEEtran}
\bibliography{bib_all_161117}{}

\end{document}